\documentclass[10pt,twocolumn]{IEEEtran}

\usepackage{adjustbox}

\usepackage{diagbox}
\usepackage{enumitem}
\usepackage{amsmath,graphics,amssymb,epsfig,subfigure,color,cite}
\usepackage{array}
\usepackage{multirow}
\usepackage{enumerate}
\usepackage{algorithmic}
\usepackage{algorithm}
\usepackage{bm}
\usepackage{float}
\usepackage{makecell}
\usepackage{lipsum}
\usepackage{capt-of}
\usepackage{amsmath}
\usepackage{amssymb}

\newtheorem{thm}{Theorem}

\newtheorem{prop}{Proposition}

\floatname{algorithm}{Algorithm}

\newcommand{\argmin}{\operatornamewithlimits{argmin}}

\begin{document}
        \title{Vision Transformer-based Semantic Communications with Importance-Aware Quantization}
	 \author{Joohyuk Park, \IEEEmembership{Student Member,~IEEE}, Yongjeong Oh, \IEEEmembership{Graduate Student Member,~IEEE}, \\ Yongjune Kim, \IEEEmembership{Member,~IEEE}, and Yo-Seb Jeon, \IEEEmembership{Member,~IEEE}
    \thanks{Joohyuk Park, Yongjeong Oh, Yongjune Kim, and Yo-Seb Jeon are with the Department of Electrical Engineering, POSTECH, Pohang, Gyeongbuk 37673, South Korea (e-mail: joohyuk.park@postech.ac.kr; yongjeongoh@postech.ac.kr; yongjune@postech.ac.kr; yoseb.jeon@postech.ac.kr).}
    }
	\vspace{-2mm}
	
	\maketitle
 
	\vspace{-12mm}

    \begin{abstract} 
        Semantic communications provide significant performance gains over traditional communications by transmitting task-relevant semantic features through wireless channels. However, most existing studies rely on end-to-end (E2E) training of neural-type encoders and decoders to ensure effective transmission of these semantic features. To enable semantic communications without relying on E2E training, this paper presents a vision transformer (ViT)-based semantic communication system with importance-aware quantization (IAQ) for wireless image transmission. The core idea of the presented system is to leverage the attention scores of a pretrained ViT model to quantify the importance levels of image patches. Based on this idea, our IAQ framework assigns different quantization bits to image patches based on their importance levels. This is achieved by formulating a weighted quantization error minimization problem, where the weight is set to be an increasing function of the attention score. Then, an optimal incremental allocation method and a low-complexity water-filling method are devised to solve the formulated problem. Our framework is further extended for realistic digital communication systems by modifying the bit allocation problem and the corresponding allocation methods based on an equivalent binary symmetric channel (BSC) model. Simulations on single-view and multi-view image classification tasks show that our IAQ framework outperforms conventional image compression methods in both error-free and realistic communication scenarios.
    \end{abstract}

    
        \section{Introduction}\label{Sec:Intro}
            Traditional communication systems primarily focus on encoding messages into bit sequences to ensure accurate reconstruction of transmitted bit sequences with minimal bit errors. However, recent advancements have shifted attention toward a more goal-oriented approach known as semantic communications \cite{SC_1, SC_2, SC_3}, which prioritizes the transmission of intended meaning over the precise recovery of bits. A typical goal of semantic communications is to maximize task performance by ensuring that the conveyed message facilitates effective task execution, even if the bit sequence is not perfectly reconstructed. This approach has gained considerable traction in resource-demanding scenarios, such as Internet of Things (IoT) wireless networks or low-latency communications, which require high data efficiency or low latency while operating under limited communication resources \cite{Application_2, Application_3}. 
            

            The most widely adopted approach in semantic communication systems is to employ joint source-channel coding (JSCC) to transmit task-related semantic features directly over wireless channels. In this approach, source and channel encoders/decoders are integrated into a single neural model, and the unified encoder and decoder models are jointly trained under wireless channel environments such as additive white Gaussian noise (AWGN) and Rayleigh fading channels. This approach has demonstrated significant performance gains over traditional separate source and channel coding methods in various applications, including image transmission \cite{image_trans_2, image_trans_3}, 
            text transmission \cite{DeepSC, ReAllo-T}, and speech transmission \cite{DeepSC-S}.   %
            Despite its success in various applications, the analog transmission assumed in the JSCC approach poses challenges for integration with existing digital communication systems, which use hardware components and processing units specifically designed for digital symbol transmission. Additionally, analog transmission has inherent drawbacks compared to digital transmission, including vulnerability to noise and limited flexibility and scalability.

            To address the limitations of the analog JSCC approach, digital semantic communication systems have been developed, focusing on efficiently representing semantic features with finite values that can be easily converted into digital symbols. For example, in \cite{DSC_Fixed_bit_2, DSC_Fixed_bit_3, DSC_Fixed_bit_5}, explicit or implicit quantization of individual semantic feature elements was employed to represent these elements as bit sequences of the same length. However, such fixed-level quantization often fails to account for feature importance, thereby limiting the system's ability to maximize task performance when communication resources are highly constrained.
            To tackle this challenge, adaptive quantization for digital semantic communications has been studied on the basis of entropy coding \cite{EC_1, EC_2} and importance-aware quantization \cite{IAQ_2,Transformer_not_att_2}. The key idea of entropy coding is to reduce the bit overhead in an average sense, by assigning shorter bit sequences to more frequent elements and longer bit sequences to less frequent ones. Based on this idea, in \cite{EC_1, EC_2}, learning-based source coders were considered to determine the optimal transmission rate of semantic features. Unlike the entropy coding which focuses on the {\em frequency} of the semantic features, the importance-aware quantization focuses on the {\em importance} of each semantic feature in the context of task performance.
            For instance, in \cite{IAQ_2}, a reinforcement learning (RL)-based bit allocation scheme was proposed for orthogonal frequency division multiplexing systems.  
            In \cite{Transformer_not_att_2}, a masking strategy was considered to mask noise-related image patches to suppress feature activations and reduce communication overhead. Both methods necessitate the use a dedicated module to compute feature importance, which needs to be carefully designed for each task.

            A common limitation of all the aforementioned studies is their reliance on end-to-end (E2E) training, where various modules in semantic communication systems are jointly trained using large amounts of training data. 
            However, the E2E training approach does not guarantee the effectiveness of the trained modules in mismatched training and testing environments, which may arise from dynamic and unpredictable wireless environments. Moreover, E2E training becomes increasingly impractical in complex scenarios such as collaborative inference, multi-modal, and multi-task settings \cite{VQA,MDCEI,MVID}. In these cases, the modules may need to be trained for all possible combinations of scenarios or, at least, re-trained for any changes in training scenarios. These requirements not only limit the scalability of these methods but also reduce their applicability in cellular or IoT sensor networks. 

            To address this limitation, semantic communication systems that do not rely on the E2E training under specific communication environments have been studied in some prior works \cite{non_e2e_add, DSC_Fixed_bit_1, DSC_Fixed_bit_4,Transformer_att_1}. 
            In \cite{non_e2e_add}, the training process involved random sampling of signal-to-noise ratio (SNR) values to ensure the JSCC model could adapt to diverse channel conditions.
            Furthermore, in \cite{DSC_Fixed_bit_1}, a digital JSCC encoder-decoder pair was trained based on a parametric-model-based training environment. Then, during inference, communication systems adapt to this training environment by adjusting modulation levels according to the parametric model. This approach significantly enhances the adaptability and flexibility of the semantic communications, but still relies on a predefined parametric model. 
            To overcome this limitation, training-free semantic communications have been suggested in \cite{DSC_Fixed_bit_4,Transformer_att_1}, which do not rely on the E2E training or the parametric-model-based training. 
            Specifically, in \cite{Transformer_att_1}, a pretrained vision transformer (ViT) encoder \cite{ViT} was used to select image patches relevant to the classification task instead of employing the E2E training. While this method effectively reduces communication overhead, the selective transmission of patches offers limited flexibility in managing communication overhead. Consequently, this approach may suffer from performance degradation when  communication resources are highly constrained (e.g., IoT sensor networks).
            More importantly, the method in \cite{Transformer_att_1} overlooks the impact of communication errors, which are inevitable due to channel fading and noise effects. This limitation restricts the broader applicability of this method in practical wireless networks.

            To take a step toward realizing training-free and practical digital semantic communications, in this paper, we present a ViT-based semantic communication system with importance-aware quantization (IAQ) for wireless image transmission. In the presented system, the importance levels of image patches are quantified using attention scores extracted from a pretrained ViT model which does not rely on the E2E training. We then adaptively assign different quantization bits to image patches based on their importance levels. This is achieved by formulating a weighted quantization error minimization problem, where the weight is defined as an increasing function of the attention score. To solve this problem, we develop two methods: (i) an optimal incremental bit allocation method and (ii) a low-complexity water-filling bit allocation method. We further adapt our framework for realistic digital communication systems by modifying the bit allocation problem and the corresponding bit allocation methods based on an equivalent binary symmetric channel (BSC) model. Simulation results on single-view and multi-view image classification tasks demonstrate the superiority of the proposed IAQ method over existing quantization approaches. The main contributions of this paper are summarized as follows:

       \begin{figure*}[t]
        \centering 
            {\epsfig{file=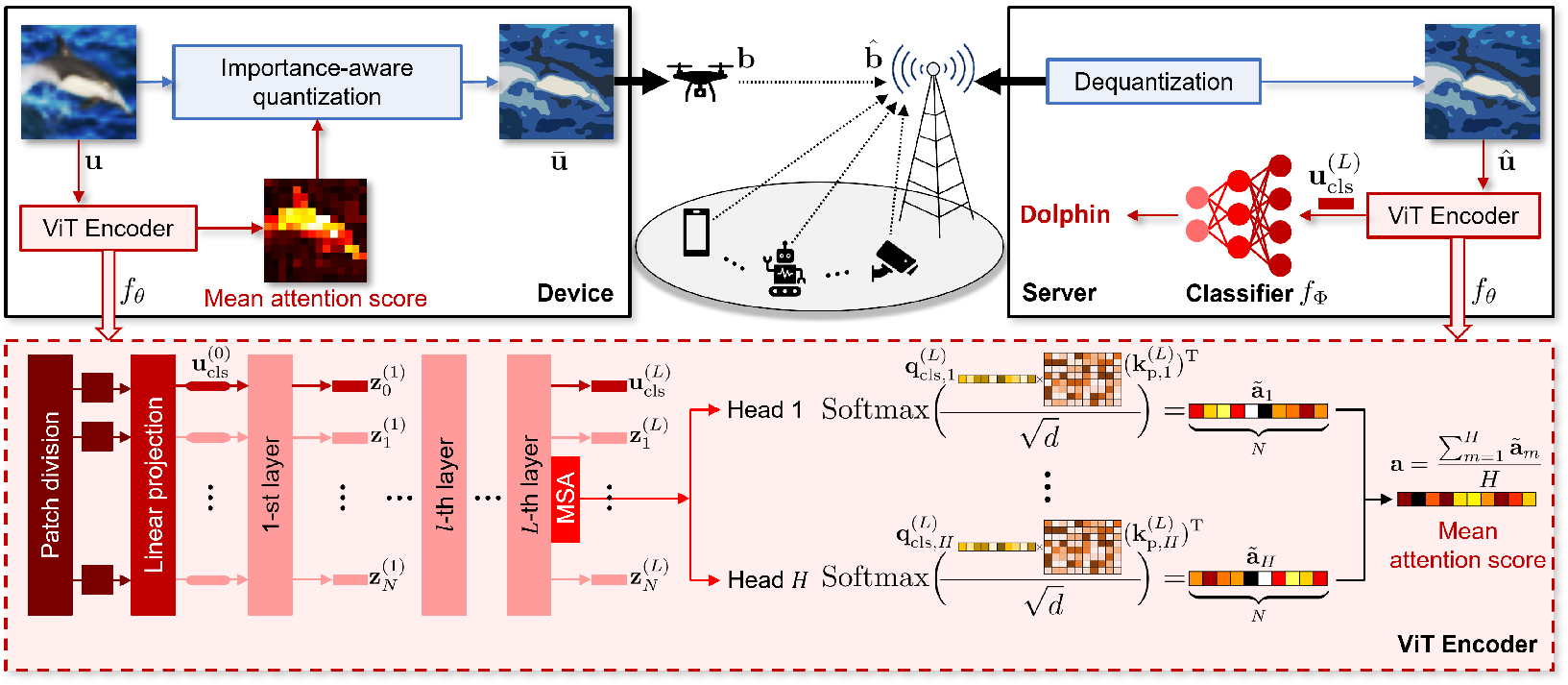, width=16cm}} 
        \caption{Illustration of the proposed IAQ framework for digital semantic communication systems.}
        \label{fig:System}
    \end{figure*}  
    
    \begin{itemize}
    \item We present a novel IAQ framework for training-free semantic communication systems. To the best of our knowledge, this is the first work that incorporates adaptive quantization into the training-free semantic communication systems. Our framework provides an enhanced flexibility in controlling communication overhead compared to a selective  transmission method considered in \cite{Transformer_att_1}. 
    
    \item We formulate an importance-aware bit allocation problem for semantic communications. To this end, we consider a weighted quantization error minimization problem by determining a weight as an increasing function of the importance level. Our problem formulation can be generalized with any choice of the weight and importance measure. Therefore, our formulation provides an optimization framework for importance-aware bit allocation that can be applied in various semantic communication systems. 

    \item We devise importance-aware bit allocation methods to solve the formulated problem. By characterizing the property of the formulated problem, we show that an incremental allocation method \cite{greedy} provides the optimal solution of the formulated problem. To reduce the computational complexity required to solve this problem, we also develop a low-complexity  method based on a simple relaxation strategy and the water-filling algorithm \cite{Water_filling_2}. 

    \item We extend our IAQ framework to operate under practical digital semantic communications, where the transmission of the bit sequence may contain errors due to channel fading and noise effects. To this end, we model the combined effects of digital communications using parallel BSCs, as done in  \cite{DSC_Fixed_bit_1}. We then reformulate the importance-aware bit allocation problem based on distortion analysis that takes into account both quantization and communication errors. We also modify the incremental allocation and water-filling methods to solve the reformulated problem. 

    \item Using simulations, we demonstrate the superiority of our IAQ framework over the existing quantization methods for single-view and multi-view image classification tasks using the CIFAR-100 \cite{CIFAR-100}, MIRO \cite{MIRO}, and MVP-N \cite{MVP_N} datasets. Our results show that the proposed framework provides significant gains in the performance-overhead tradeoff for the considered tasks, compared to the existing approaches.

    \end{itemize}
    


    \section{System Model}\label{Sec:Model}
    In this work, we consider a ViT-based semantic communication system for wireless image transmission, where a device transmits an image to a server performing a dedicated machine learning task (e.g., image classification). In our system, the device employs a lightweight ViT encoder, and the server uses a more complex one, reflecting the resource constraints of IoT devices and the computational capacity of servers \cite{Transformer_att_1}.
    We assume that both the encoders are pre-trained using large datasets.


     The ViT encoder equipped at the device is denoted by the function $f_{\theta}(\cdot)$, parameterized by weights $\theta$. Given input data ${\bf{u}} \in \mathbb{R}^{H\times W\times C}$, where $H$, $W$, and $C$ represent the height, width, and number of channels, respectively, we first reshape ${\bf u}$ into a sequence of flattened 2D patches, denoted by ${\bf u}_{\rm p} \in \mathbb{R}^{N\times P^2C}$. Here, $(P,P)$ is the patch size, and $N=\frac{HW}{P^2}$ is the total number of patches. By using a projection matrix ${\bf T} \in \mathbb{R}^{P^2C\times D}$, these patches are linearly transformed to a $D$-dimensional vectors, i.e., 
    \begin{align}\label{eq:embed}
        {\bf z}^{(0)} = \big[{\bf u}^{(0)}_{\rm cls}; {\bf u}_{{\rm p}, 1}{\bf T}; \ldots;  {\bf u}_{{\rm p}, N}{\bf T}\big]+{\bf T}_{\rm pos} \in \mathbb{R}^{(N+1)\times D},
    \end{align}
    where ${\bf u}_{{\rm p},i}\in\mathbb{R}^{1\times P^2C}, i \in \{1,\cdots,N\}$ is the $i$-th row of ${\bf u}_p$, ${\bf z}^{(0)}$ is the output of linear projection, and ${\bf T}_{\rm pos}$ denotes the learnable position embedding. 
    To facilitate the classification process, a class token ${\bf z}_0^{(0)} = {\bf u}^{(0)}_{\rm cls} \in \mathbb{R}^{1 \times D}$ is added at the beginning of the embedded patch sequence. This class token is crucial for gathering information from the entire sequence, ultimately contributing to the final classification result.
    The overall forwarding mechanism of the ViT encoder, described in \cite{ViT}, can be summarized as
    \begin{align}\label{eq:encoder}
        {\tilde{\bf z}}^{(l)}_j &= {\rm MSA}({\rm LN}({\bf z}^{(l-1)}_j))+{\bf z}^{(l-1)}_j, \nonumber\\
        {\bf z}^{(l)}_j &= {\rm MLP}({\rm LN}({\tilde{\bf z}}^{(l)}_j))+{\tilde{\bf z}^{(l)}_j}, \\
        {\bf u}^{(L)}_{\rm cls} &= {\rm LN}({\bf z}^{(L)}_0), \nonumber
    \end{align}
    where $j \in \{0, 1, \ldots, N\}$, $l \in \{1, \ldots, L\}$, and $L$ is the number of layers. $\rm MSA$, $\rm MLP$ and $\rm LN$ represent the multi-head self-attention, multi-layer perceptron and layer normalization, respectively. The final class token ${\bf u}^{(L)}_{\rm cls}$ is obtained from the encoder's output and then used as input to the classifier, defined by the function $f_{\Phi}(\cdot)$ with parameters $\Phi$.

    To determine the significance of each patch, we leverage the attention scores produced by the MSA mechanism. The $n$-th single-head attention score $\tilde{\bf a}_m$ is calculated as
    \begin{align}\label{eq:attention}
        \tilde{\bf a}_m = {\rm Softmax}\left(\frac{{\bf q}^{(L)}_{{\rm cls}, m}({\bf k}_{{\rm p}, m}^{(L)})^{\rm T}}{\sqrt{d}}\right) \in \mathbb{R}^{1\times N},
    \end{align}
    where $m \in \{1,\cdots,H\}$ and $H$ is the total number of heads. ${\bf q}^{(L)}_{{\rm cls},m} \in \mathbb{R}^{1\times d}$ represents the query for the class token at the $m$-th head and the $L$-th layer, while ${\bf k}^{(L)}_{{\rm p},m} \in \mathbb{R}^{N\times d}$ denotes the keys corresponding to the image patches in the same head and layer \cite{non_e2e_add, ViT}. Finally, the mean attention score ${\bf a} \in \mathbb{R}^{1\times N}$ is obtained as
    \begin{align}\label{eq:attention2}
        {\bf a} = \frac{\sum_{m=1}^{H}\tilde{\bf a}_m}{H}.
    \end{align}

    After extracting the mean attention scores, these scores are utilized as an input for a  quantization process applied to the input image $\bf u$. This process yields a quantized image $\bar{\bf u}$ and the corresponding bit sequence ${\bf b} \in \{0,1\}^{B}$, where $B=P^2C\sum_{i=1}^{N}M_i$ represents the total length of bit sequence and $M_i$ denotes the quantization bit for the $i$-th patch. Details of the quantization method will be introduced in Sec.~\ref{Sec:IAQ_error_free} and Sec.~\ref{Sec:IAQ_error}.
    The process of transmitting the bit sequence $\bf b$ to the server using digital communications can be equivalently modeled using BSCs, as will be explained in Sec.~\ref{Sec:BSC}. Then, at the server, the received bit sequence $\hat{\bf b}$ is transformed into a reconstructed image $\hat{\bf u}$ through a dequantization process. After reconstructing the image $\hat{\bf u}$, the ViT encoder at the server is applied to perform the dedicated machine learning task (e.g., image classification task). The overall transmission and reception process considered in our work is illustrated in Fig.~\ref{fig:System}.



    \begin{figure}
    \centering
    \begin{minipage}{\columnwidth}
        \centering
        \subfigure[Error-free communication]
        {\epsfig{file=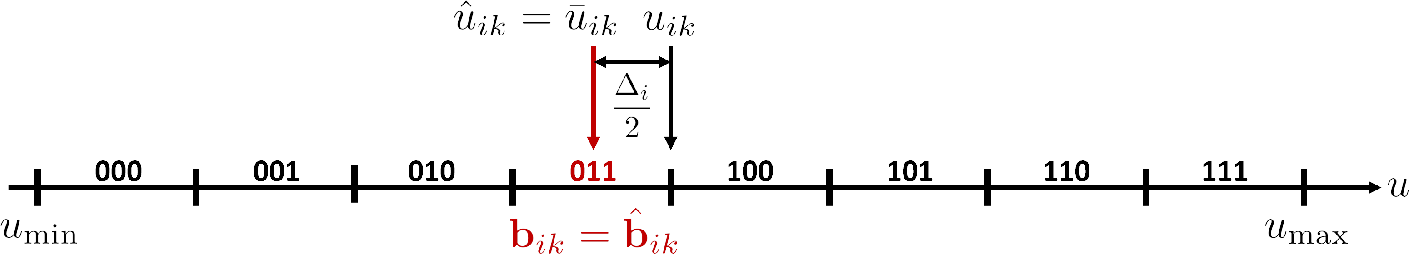, width=\linewidth}}
    \end{minipage}
    \begin{minipage}{\columnwidth}
        \centering
        \subfigure[Communication with one-bit error]
        {\epsfig{file=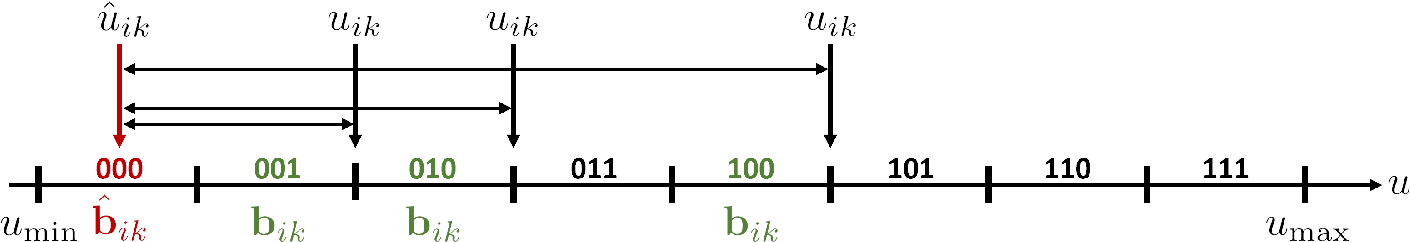, width=\linewidth}}
    \end{minipage}
    \caption{Illustration of the operation of a uniform quantizer under error-free communication and communication with one-bit error.}
    \label{fig:UQ}  
    \end{figure}

    \section{Proposed IAQ Framework for VIT-based Semantic Communications}\label{Sec:IAQ_error_free}
    In this section, we present a novel IAQ framework to enable an efficient image transmission for the digital semantic communication system. We begin by addressing a simple case under an error-free transmission scenario, where the BSC parameter $\tilde{\mu}$ is set to 0. An extension to an erroneous transmission scenario will be discussed in Sec.~\ref{Sec:IAQ_error}.

    \subsection{Patch-Wise Quantization}\label{Sec:IAQ_error_free1}
    We start by explaining a patch-wise quantization approach adopted in our framework. Let $u_{ik}$ be the $k$-th pixel value of the $i$-th patch of the original image $\mathbf{u}$, where $i \in \{1,\ldots,N\}$ and $k \in \{1,\ldots,P^2C\}$. 
    In this work, we simply adopt a uniform quantizer which is the most representative scalar quantization technique. 
    If the uniform quantizer with the quantization bit $M_i \geq 0$ is assigned to the $i$-th patch, the quantizer output for ${u}_{ik}$ is expressed as
     \begin{align}\label{eq:uni_quant}
        \bar{u}_{ik} = \argmin_{q \in \mathcal{Q}_{i}}~|q - {u}_{ik}|^2,
    \end{align}
    where $\mathcal{Q}_{M_i} = \big\{ u_{\rm min} + (s+1/2)\Delta_i | s\in  \{0, \ldots, 2^{M_i}-1\}  \big\}$, $\Delta_i = (u_{\rm max}-u_{\rm min})/2^{M_i}$, $u_{\rm min} = \underset{i,k}{\rm min}(u_{ik})$, and $u_{\rm max} = \underset{i,k}{\rm max}(u_{ik})$ denote the minimum and maximum pixel values of an 8-bit image, respectively.
    %
    Once $\bar{u}_{ik}$ is obtained, it can be equivalently represented using a binary sequence ${\bf b}_{ik} = \big[b_{ik,1},\ldots,b_{ik,M_i}\big]$, given by 
    \begin{align}\label{eq:uni_quant_bit}
        {\bf b}_{ik}=\bigg[\frac{\bar{u}_{ik}-u_{\rm min}}{\Delta_i}-\frac{1}{2}\bigg]^{(2)},
    \end{align}
    where the operator $[\cdot]^{(2)}$ denotes the binary representation of an integer. 
    An example of the uniform quantizer with $M_i=3$ under error-free communications is shown in Fig.~\ref{fig:UQ}(a). In this scenario, if $u_{\rm min}+3\Delta_i < u_{ik} < u_{\rm min}+4\Delta_i$, we have $\bar{u}_{ik}=u_{\rm min}+3.5\Delta_i$, and the corresponding bit sequence is given by ${\bf b}_{ik} = \big[3\big]^{(2)}=011$. 
    


    We now analyze the quantization error of our patch-wise quantization approach. For the $i$-th patch, the quantization error of each pixel value is upper bounded by the maximum difference between the quantization output $\bar{u}_{ik}$ and the input $u_{ik}$, as depicted in Fig.~\ref{fig:UQ}(a). Therefore, the upper bound of the quantization error between $\bar{u}_{ik}$ and the input $u_{ik}$ is given by 
    \begin{align}\label{eq:UB_ED_error_free_1}
        |u_{ik}-\bar{u}_{ik}|^2 \leq \left(\frac{\Delta_i}{2}\right)^2 =  \frac{(u_{{\rm max}}-u_{{\rm min}})^2}{4 } 4^{-M_i},
    \end{align}
    for all $i$ and $k$. 
    Consequently, the upper bound of the quantization error of the $i$-th patch is expressed as
    \begin{align}\label{eq:UB_ED_error_free_2}
        \left\|{\bf u}_i-\bar{\bf u}_i \right\|^2 &= \sum_{k=1}^{P^2C} |u_{ik}-\bar{u}_{ik}|^2 \nonumber \\
        &\leq  \underbrace{P^2 C \frac{(u_{{\rm max}}-u_{{\rm min}})^2}{4 }}_{\triangleq D_0} 4^{-M_i}.
    \end{align}
    As shown in \eqref{eq:UB_ED_error_free_2}, the quantization error of the $i$-th patch exponentially decreases with the assigned quantization bit $M_i$.
    

    After the patch-wise quantization, a complete bit sequence ${\bf b}$ is constructed by concatenating each ${\bf b}_{ik}$ across all $i$ and $k$. 
    Since we focus on the error-free transmission of the bit sequence ${\bf b}$ in this section, the received sequence $\hat{\bf b}$ at the server is identical to the transmitted sequence. Otherwise, as illustrated in Fig.~\ref{fig:UQ}(b), the received sequence differs from ${\bf b}$.
    The received bit sequence is divided into patch-wise sequences, denoted by $\{\hat{\bf b}_{ik}\}_{\forall i,k}$. Each patch-wise sequence $\hat{\bf b}_{ik}$ then undergoes dequantization to produce the $k$-th pixel value of the $i$-th patch, denoted by $\hat{u}_{ik}$. Finally, a dequantized image $\hat{\bf u}$ is constructed by rearranging the dequantized values into an image.


    \begin{figure}[t]
        \centering 
            {\epsfig{file=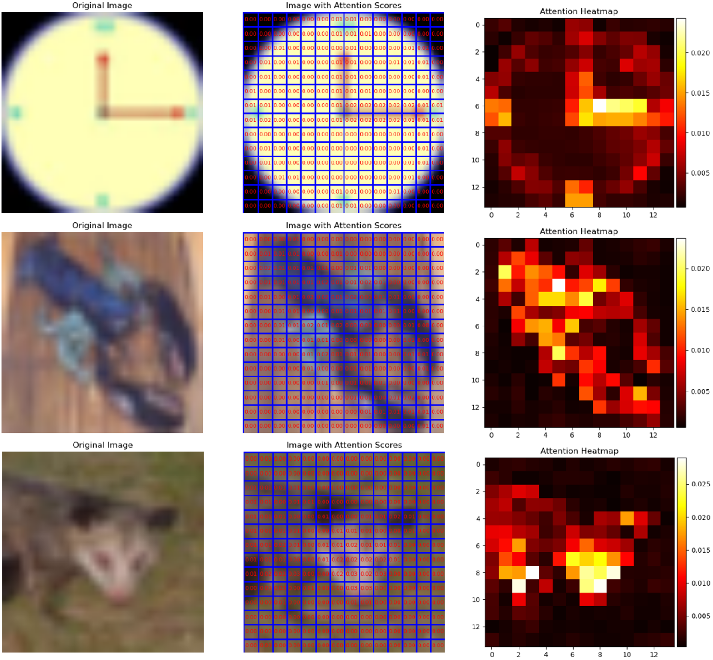, width=6.9cm}}
        \caption{Visualization of the attention score map measured using ViT encoder for the single-view image classification task on the CIFAR-100 dataset.}
        \label{fig:ATT}
    \end{figure}
    
    \subsection{Importance-Aware Bit Allocation: Incremental Allocation}\label{Sec:IAQ_error_free2}
    Typically, each patch within the image exhibits significantly different mean attention scores, indicating that the patches have varying impacts on task performance \cite{ATT_diff}. For example, Fig.~\ref{fig:ATT} visualizes the attention score map generated by the ViT encoder for an image classification task on the CIFAR-100 dataset. This figure shows that the attention scores are relatively high for critical pixels in the images, while they are relatively low for less important pixels. Inspired by this observation, in our IAQ framework, we adopt mean attention scores as an operational {\em measure} of the importance level of each patch within the image. We then allocate varying quantization bits to patches based on their mean attention scores, assigning higher quantization bits to patches with higher mean attention scores.


    Based on this idea, we formulate a weighted quantization error minimization problem by leveraging the mean attention score of the patch as a weight for the quantization error.
    By utilizing the upper bound of the quantization error in \eqref{eq:UB_ED_error_free_2}, our problem is formulated as 
    \begin{align}
        ({\bf P}_1)~~\underset{\{M_i\}^{N}_{i=1}}{\rm min}~ &\sum_{i=1}^{N} w(a_i) D_0 4^{-M_i}, \label{eq:opt_DMO_1}\\
        {\rm s.t.}~~~ &  B + B_{\rm add} \leq B_{\rm target}, \label{eq:opt_DMO_1_c1}\\
        &M_{i} \in \{0,\ldots,M_{\rm max}\}, \forall i \in \left\{1, \ldots, N\right\}, \label{eq:opt_DMO_1_c2}
    \end{align}
    where $B_{\rm target}$ is a total bit constraint such that $P^2CNM_{\rm max} + B_{\rm quant} > B_{\rm target}$, $M_{\rm max}$ is the maximum quantization bit of the uniform quantizer, and $w(a_i)$ is a {\em weight} function which can be any arbitrary increasing function of $a_i$. Here, $B_{\rm add}$ is the bit overhead required to transmit the quantizer information, defined as
    \begin{align}\label{eq:ex_bit_overhead}
       B_{\rm add} =  \big\lceil \log_2(M_{\rm max}+1) \big \rceil  N+8\cdot2,
    \end{align}
    where $\left \lceil \cdot \right \rceil$ is the ceiling function. The first term represents the bit overhead for $\left\{M_i\right\}_{i=1}^{N}$, and the second term denotes the bit overhead for transmitting $u_{\rm max}$ and $u_{\rm min}$. This quantizer information enables the server to perform dequantization. It should be noted that, in most cases, this additional overhead is negligible compared to the bit overhead of $B$ required for transmitting the quantized pixel values of the image.

    The problem $({\bf P}_1)$ is categorized as a discrete optimization problem. Since the objective function of $({\bf P}_1)$ is convex and monotonically decreasing function of $M_i$, it is well known that the optimal solution of $({\bf P}_1)$ is obtained via an incremental allocation algorithm \cite{greedy}. This algorithm begins by initializing $M_i = 0$, $\forall i$, to start from the minimal bit overhead. Then, for each patch where $M_i$ has not yet reached $M_{\rm max}$, the algorithm iteratively increases $M_i$ by 1 for the patch that leads to the largest decrease in the objective function as a response to the additionally assigned bit. This process continues until the total bit constraint in \eqref{eq:opt_DMO_1_c1} is satisfied, progressively increasing the quantization bit for the selected patch. This algorithm offers an optimal allocation of the quantization bits. 

    
    \subsection{Importance-Aware Bit Allocation: Water Filling}\label{Sec:IAQ_error_free3}
    A major limitation of the incremental allocation method in Sec.~\ref{Sec:IAQ_error_free2} is its substantial computational complexity, quantified as $\mathcal{O}\left(N \frac{\bar{B}}{P^2 C}\right)$, where $\bar{B} = B_{\rm target} - B_{\rm add}$. In each iteration, the objective function given in \eqref{eq:opt_DMO_1} must be computed, resulting in a complexity of $\mathcal{O}(N)$. This is further multiplied by the total number of iterations, $\frac{\bar{B}}{P^2 C}$.
    To tackle this limitation, we develop a low-complexity bit allocation method based on the water-filling algorithm. 
    To this end, we reformulate the problem $({\bf P}_1)$ by substituting the quantization bit $M_i$  with the quantization level $Q_i = 2^{M_i}$ and also by relaxing the quantization level $Q_i$ from discrete values to continuous real values. Our reformulated problem is given by 
    \begin{align}
        ({\bf P}_2)~~\underset{\{Q_i\}^{N}_{i=1}}{\rm min}~ &\sum_{i=1}^{N}w(a_i)  D_0 Q_i^{-2}, \label{eq:opt_DM_1}\\
        {\rm s.t.}~~~ & B + B_{\rm add} \leq B_{\rm target}, \label{eq:opt_DM_1_c1}\\
        &1 \leq Q_{i} \leq Q_{\rm max}, \forall i \in \left\{1, \ldots, N\right\}, \label{eq:opt_DM_1_c2}
    \end{align}
    where $Q_{\rm max} = 2^{M_{\rm max}}$. The problem $({\bf P}_2)$ is categorized as a convex optimization problem, specifically a cave-filling problem within the family of water-filling problems \cite{Water_filling_2}. Consequently, the optimal solution of $({\bf P}_2)$ can be obtained by applying the Karush-Kuhn-Tucker (KKT) conditions.

    The optimal quantization level $Q^\star_i$ for the $i$-th patch is given in the following theorem:

    \begin{thm}
            The optimal solution of the problem $({\bf P}_2)$ is
    \begin{align}\label{eq:opt_DM_5}
        Q^\star_i =\! {\rm min}\!\left\{\!Q_{\rm max}, {\rm max}\left\{1, \sqrt{\frac{w(a_i)  \ln 2(u_{\rm max}-u_{\rm min})^2}{2\nu^\star}}\right\}\!\right\},
    \end{align}
    for all $i \in \{1,\ldots, N\}$, where $\nu^\star$ is the optimal Lagrange multiplier determined to satisfy the following equality:
    \begin{align}
        \sum_{i=1}^{N}\log_2 Q^\star_i  = \frac{\bar{B}}{P^2C}.
    \end{align}
    \end{thm}
    \begin{IEEEproof}
        See Appendix A.
    \end{IEEEproof}
    
    \vspace{2mm}
    The optimal Lagrange multiplier $\nu^\star$ in Theorem 1 can be easily obtained using various water-filling algorithms, such as the bisection search algorithm \cite{Water_filling_2} and the fast water-filling algorithm \cite{Water_filling_3}. Once $\nu^\star$  is acquired through the water-filling algorithm, the optimal quantization level is determined as shown in \eqref{eq:opt_DM_5}. 
    
    Theorem 1 demonstrates that the optimal quantization level increases with the weight $w(a_i)$ which is the increasing function of the mean attention score. Therefore, the quantization level allocation in \eqref{eq:opt_DM_5} assigns a higher quantization level to patches with higher mean attention scores compared to those with lower scores. This indicates that our quantization framework with the optimal bit allocation effectively reduces overall quantization error by adaptively quantizing patches according to their importance.
    

    It should be noted that the optimal quantization level in \eqref{eq:opt_DM_5} may not be a power of two, which violates the assumption of $Q_i = 2^{M_i}$ in our work. To address this issue, we introduce a simple adjustment step to satisfy the above assumption while utilizing the entire available bit budget. This step involves rounding the solutions in \eqref{eq:opt_DM_5} (i.e., ${\sf round off}(\log_2 Q^\star_i) = \hat{m}_i$), and then adjusting the quantization levels based on $B_{\rm target}$ and the differences denoted by $D_{\rm bit} = \bar{B} -P^2C\sum_{i=1}^{N}\hat{m}_i$. 
    Then, if $D_{\rm bit} > 0$ and $2^{\hat{m}_i} < Q_{\rm max}$, one bit is incrementally added to the patch (i.e., $\hat{Q}_i = 2^{\hat{m}_i+1}$), with this adjustment performed in descending order of mean attention scores. Conversely, if $D_{\rm bit} < 0$ and $2^{\hat{m}_i} > 1$, one bit is incrementally removed from the patch (i.e., $\hat{Q}_i = 2^{\hat{m}_i-1}$), with this adjustment carried out in ascending order of mean attention scores. This step ensures that the total bit constraint in \eqref{eq:opt_DM_1_c1} is satisfied with the adjusted quantization bits.

    \vspace{1mm}
    {\bf Remark 1 (Complexity comparison of the bit allocation methods):}
    The problem $({\bf P}_1)$ has the same objective function and bit overhead constraint as the problem $({\bf P}_2)$, with the distinction that $({\bf P}_1)$ assumes discrete quantization levels while $({\bf P}_2)$ considers continuous ones. 
    As a result, \eqref{eq:opt_DM_1_c2} represents a relaxed feasible set of \eqref{eq:opt_DMO_1_c2}, implying that the optimal solution of $({\bf P}_2)$ is suboptimal for the original problem $({\bf P}_1)$.
    The computational complexity order required to solve $({\bf P}_2)$ is $\mathcal{O}(NS_{\rm max})$, where $S_{\rm max}$ denotes the maximum number of iterations required by the water-filling algorithm.
    In contrast, solving $({\bf P}_1)$ requires a computational complexity of $\mathcal{O}(N\frac{\bar{B}}{P^2 C})$.
    Since $S_{\rm max} \leq N \leq \frac{\bar{B}}{P^2 C}$ typically holds due to the rapid convergence of the water-filling algorithm and the required minimum bit overhead, $({\bf P}_2)$ is computationally more efficient than $({\bf P}_1)$. 
    Therefore, the water-filling algorithm demonstrates advantages in terms of computational complexity compared to the incremental allocation method.

    \section{Extension to Realistic Digital Semantic Communications}\label{Sec:IAQ_error}
    In this section, we extend our IAQ framework to operate under practical digital semantic communications, where the transmission of the bit sequence suffers from communication errors due to channel fading and noise effects. 

    \subsection{BSC Modeling Approach}\label{Sec:BSC}

    We start by describing a standard process to incorporate our IAQ framework into practical digital semantic communication systems. In this process, the bit sequence ${\bf b}$, corresponding to the quantized image $\bar{\bf u}$, is modulated into digital symbols ${\bf x} = \big[x[1], \cdots, x[T]\big]^{\sf T} \in \mathcal{X}^T$, where $\mathcal{X}$ denotes the constellation set, and transmitted through fading channels.
    Then a received signal at time slot $t$ is given by 
     \begin{align}
        y[t] = h[t]x[t] + v[t],
    \end{align}
    where $h[t] \in \mathbb{C}$ represents a channel gain at time slot $t$, and $v[t] \in \mathbb{C}$ is an AWGN distributed as $\mathcal{CN}(0, \sigma^2)$.
    At the server, channel estimation is performed to obtain an estimate of $h[t]$. Based on this, channel equalization is applied to compensate for the effects of fading, commonly using a zero-forcing (ZF) equalizer \cite{BER}. The equalized signal is then passed through data detection, which determines the transmitted symbol $\bf x$ from  received signals. Following the data detection, symbol demapping is employed to recover the estimated bit sequence $\hat{\bf b}$. Then this bit sequence $\hat{\bf b}$ is applied as an input of the dequantization process, yielding a reconstructed image $\hat{\bf u}$.

    Our key strategy is to equivalently model the relationship between the input ${\bf b}$ and output $\hat{\bf b}$ of the digital communication process by using parallel BSCs with certain bit-flip probabilities, as explained in \cite{DSC_Fixed_bit_1}. 
    Specifically, these BSCs have the same  bit-flip probabilities when the same modulation is applied to all the bits in ${\bf b}$, and the channel remains constant within the transmission of the symbol sequence ${\bf x}$ (i.e., $h[t] = h$, $\forall t$) \cite{BER}.
    Then, the relationship between ${\bf b}$ and $\hat{\bf b}$ are equivalent to $B$-parallel BSCs with a common bit-flip probability $\tilde{\mu} \in [0,1]$.  
    This implies that the conditional distribution of $\hat{b}_{j}$ for a given ${b}_{j}$ is represented as 
    \begin{align}\label{channel_model}
        p_{\mathrm{BSC}}(\hat{b}_{j}|b_{j};\tilde{\mu}) = 
        \begin{cases}
            1-\tilde{\mu}, & \text{if}~\hat{b}_{j} = b_{j}, \\
            \tilde{\mu}, & \text{if}~\hat{b}_{j} \neq b_{j}. 
        \end{cases}
    \end{align}

    A key advantage of our BSC modeling approach is that it can reflect various communication environments without explicitly considering digital modulation, fading channel, channel equalization, and digital demodulation processes; rather, the combined effect of these processes is implicitly captured by the parallel BSCs with a bit-flip probability  $\tilde{\mu}$. Note that $\tilde{\mu}$ is equivalent to the bit error rate (BER) performance of the system and therefore characterized as a function of the channel $h$, the SNR $=\mathbb{E}[|x[t]|^2] / \mathbb{E}[|v[t]|^2]$, and the constellation set $\mathcal{X}$ \cite{BER}. 
    
    In practical communication scenarios where bit-flip error is inevitable (i.e., $\tilde{\mu} > 0$), the errors between ${\bf b}$ and $\hat{\bf b}$ occur before the dequantization process, leading to a degradation in task performance. This motivates us to devise a new quantization technique that minimizes the distortion between the input image ${\bf u}$ and the reconstructed image $\hat{\bf u}$, taking into account both quantization and communication errors.

    \subsection{Distortion Analysis}\label{Sec:Distortion}
    We analyze the distortion of our patch-wise quantization approach, by taking into account both quantization and communication errors. 
    A key observation behind our analysis is that the length of a {\em pixel-wise} bit sequence ${\bf b}_{ik}$, given by $M_i$, is very short because the maximum quantization bit is  a small number (e.g., $M_{\rm max} = 8$) in general. Additionally, practical communication systems often operate in a small bit-error-rate regime (i.e., $\tilde{\mu} \ll 1$). These facts imply that the probability of having more than two-bit error within each pixel-wise sequence is very small and therefore negligible. Motivated by this, we focus on characterizing the distortion under the following assumption: 
    
    \vspace{2mm}
    {\bf Assumption 1:} The maximum number of the bit-flip errors occurred within each pixel-wise bit sequence ${\bf b}_{ik} \in [0,1]^{M_i}$ is limited to one, i.e., $\|{\bf b}_{ik} - \hat{\bf b}_{ik}\|_0  \leq 1$, $\forall i,k$. 
    \vspace{2mm}
    
    Based on {\bf Assumption 1}, we characterize the upper bound of the expected distortion for the $i$-th patch. This result is stated in the following theorem.
    
    \vspace{1mm}
    \begin{thm}
         The upper bound of the expected distortion for the $i$-th patch under {\bf Assumption 1} is    
         \begin{align}\label{eq:one_approx_1}
         \sum_{k=1}^{P^2C}\mathbb{E}_{{\bf b}_{ik},\hat{\bf b}_{ik}}[|u_{ik}-\hat{u}_{ik}|^2] \leq D(Q_i;\tilde{\mu}),
    \end{align}
    for all $i \in \{1,\ldots, N\}$, where 
    \begin{align}\label{eq:one_approx_2}
         D(Q_i;\tilde{\mu}) = \frac{D_0Q_{i}^{\log_2 (\frac{1-\tilde{\mu}}{4})}}{(1-\tilde{\mu})} \left\{\frac{4}{3}\tilde{\mu} Q_i^2+4 \tilde{\mu} Q_i + \left(1-\frac{16}{3}\tilde{\mu} \right) \right\}.
    \end{align}
    \end{thm}
    \begin{IEEEproof}
        See Appendix B. 
    \end{IEEEproof}
    \vspace{1mm}
    
    
    If $\tilde{\mu} = 0$ in \eqref{eq:one_approx_1}, it becomes equivalent to \eqref{eq:UB_ED_error_free_2}. Therefore, \eqref{eq:one_approx_1} represents a generalized upper bound that encompasses both quantization and communication errors.

    \subsection{Importance-Aware Bit Allocation for Distortion Minimization}\label{Sec:IAQ_error3}
    On the basis of the distortion analysis in Sec.~\ref{Sec:Distortion}, we modify the importance-aware bit allocation methods in Sec.~\ref{Sec:IAQ_error_free2} and Sec.~\ref{Sec:IAQ_error_free3} to incorporate them into the realistic digital semantic communications. 

    \subsubsection{Modified incremental allocation}
    In this modification, we formulate a weighted distortion minimization problem by utilizing the distortion upper bound in \eqref{eq:one_approx_1} along with the weight function $w(a_i)$. Our bit allocation problem is given by 
    \begin{align}
        \label{eq:OBEA_DMO_21}
        ({\bf P}_3)~~\underset{\{M_i\}^{N}_{i=1}}{\rm min}&\sum_{i=1}^{N}w(a_i) D(2^{M_i}; \tilde{\mu}), \\
        \label{eq:OBEA_DMO_22}
        {\rm s.t.}~~& B + B_{\rm add} \leq B_{\rm target},\\
        \label{eq:OBEA_DMO_23}
        &M_{i} \in \{0,\ldots,M_{\rm max}\}, \forall i \in \left\{1, \ldots, N\right\}.
    \end{align}

    To solve the problem  $({\bf P}_3)$, we characterize the property of  the distortion function $D(2^{M_i}; \tilde{\mu})$ with respect to $M_i$, as given in the following proposition:

    \vspace{1mm}
    \begin{prop}
        If $\tilde{\mu} \leq \frac{3}{13}$, the distortion function $D(2^{M_i}; \tilde{\mu})$ is convex and monotonically decreasing with respect to $M_i$ for $M_i \geq 0$.
    \end{prop}
    \begin{IEEEproof}
        See Appendix C.
    \end{IEEEproof}
    \vspace{1mm}
    

    Proposition 1 indicates that the problem $({\bf P}_3)$ has the form of a well-known discrete optimization problem and therefore can be solved using an incremental allocation algorithm \cite{greedy}. For this reason, we refer to the above bit allocation method as a modified incremental allocation method. Note that if there is no communication error, our modified  method is identical to the method in Sec.~\ref{Sec:IAQ_error_free2} because the objective function of the problem $({\bf P}_3)$ with $\tilde{\mu} = 0$ is the same as that of $({\bf P}_1)$, (i.e., $D(2^{M_i}; \tilde{\mu}=0)= D_0 4^{-M_i}$).

    \vspace{2mm}
    \subsubsection{Modified water filling} 
    In this modification, we reformulate the weighted distortion minimization problem $({\bf P}_3)$ by substituting the quantization bit with a real-valued quantization level, as done in Sec.~\ref{Sec:IAQ_error_free3}.  The reformulated problem is given by 
    \begin{align}
        \label{eq:OBEA_DM_21}
        ({\bf P}_4)~~~\underset{\{Q_i\}^{N}_{i=1}}{\rm min}&\sum_{i=1}^{N}w(a_i)D(Q_i; \tilde{\mu}), \\
        \label{eq:OBEA_DM_22}
        {\rm s.t.}~~ &B+ B_{\rm add} \leq B_{\rm target},\\
        \label{eq:OBEA_DM_23}
        &1 \leq Q_{i} \leq Q_{\rm max}, \forall i \in \left\{1, \ldots, N\right\}.
    \end{align}
    To solve the problem  $({\bf P}_4)$, we characterize the property of the distortion function $D(Q_i; \tilde{\mu})$ with respect to $Q_i$. The result is stated in the following proposition:

    \vspace{1mm}
    \begin{prop}
        If $\tilde{\mu} \leq \frac{3}{13}$, the distortion function $D(Q_i; \tilde{\mu})$ is convex with respect to $Q_i$ for $Q_i \geq 1$.
    \end{prop}
    \begin{IEEEproof}
        See Appendix D. 
    \end{IEEEproof}
    \vspace{1mm}

    Since a positive weighted sum of convex functions is also convex, the problem (${\bf P}_4$) is a convex optimization problem. By utilizing the KKT conditions, we derive the optimal solution of the problem $({\bf P}_4)$ as given in the following theorem:

    \vspace{1mm}
    \begin{thm}
        If $\tilde{\mu} < \frac{3}{13}$, the optimal solution of the problem $({\bf P}_4)$ is
    \begin{align}\label{eq:OBEA_DM_3}
        Q^\star_i = {\rm min}\left\{Q_{\rm max}, {\rm max}\left\{1, h^{-1} (\nu^\star;w(a_i), \tilde{\mu}) \right\}\right\},
    \end{align}
    for all $i \in \{1,\ldots, N\}$, where $h^{-1} (\cdot ;w(a_i), \tilde{\mu})$ is the inverse function of $h(\cdot ;w(a_i), \tilde{\mu})$ defined in \eqref{eq:h_def}, and $\nu^\star$  is the optimal Lagrange multiplier determined to satisfy the following equality:
    \begin{align}
        \sum_{i=1}^{N}\log_2 Q^\star_i  = \frac{\bar{B}}{P^2C}.
    \end{align}
    \end{thm}
    \begin{IEEEproof}
         See Appendix A. 
    \end{IEEEproof}
    \vspace{1mm}
    
    \begin{figure*}
    \begin{align}\label{eq:h_def}
        h(Q_i; w(a_i), \tilde{\mu}) \triangleq \frac{w (a_i) D_0 \ln 2}{P^2C(1-\tilde{\mu})}Q_i^{\log_2 (\frac{1-\tilde{\mu}}{4})}\left\{-\frac{4}{3}\tilde{\mu}Q_i^{2}\log_2 (1-\tilde{\mu})-4\tilde{\mu}Q_i\log_2 \left(\frac{1-\tilde{\mu}}{2}\right)-\left(1-\frac{16}{3}\tilde{\mu}\right)\log_2 \left(\frac{1-\tilde{\mu}}{4}\right)\right\}.
    \end{align}
    \hrulefill	
    \end{figure*}  
    \begin{algorithm}[t]
        \caption{Modified Water Filling}\label{alg:WDM}
    	{\small
    	{\begin{algorithmic}[1]
            \REQUIRE $\{w(a_i)\}_{i=1}^{N}, \tilde{\mu}$, $B_{\rm target}$
            
            \ENSURE Optimal quantization level $\{Q^\star_i\}^N_{i=1}$
            \STATE Set $\nu_{\rm min}$ and $\nu_{\rm max}$ to be sufficiently small and large values, respectively.
            \STATE $Q^{(0)}_i=1, \forall i \in \{1,\ldots, N\}$
            \FOR {$s=1$ to $S_{\rm max}$}
            \STATE $\nu = (\nu_{\rm min}+\nu_{\rm max})/2$
            \FOR {$t=0$ to $T_{\rm max}$}
            \STATE $\tilde{h}(Q^{(t)}_i; w(a_i), \tilde{\mu}) = h(Q^{(t)}_i; w(a_i), \tilde{\mu})-\nu, \forall i$
            \STATE $Q^{(t+1)}_{i} = Q^{(t)}_i-\tilde{h}(Q^{(t)}_i; w(a_i), \tilde{\mu})/\tilde{h}'(Q^{(t)}_i; w(a_i), \tilde{\mu}), \forall i$
            \STATE {\bf if} {$|Q_i^{(t+1)}-Q_i^{(t)}| < \tau_q $} {\bf then}
            \STATE ~~${Q}^\star_i = \min((\max(Q^{(t)}_i, 1), Q_{\rm max}), \forall i$
            \STATE ~~Break the loop
            \STATE {\bf end}
            \ENDFOR
            \STATE $B = P^2C\sum_{i=1}^{N}\log_2 {Q}^\star_i$
            \STATE {\bf if} {$B < \bar{B}$} {\bf then}
            \STATE ~~$\nu_{\rm max} = \nu$
            \STATE {\bf else}
            \STATE ~~$\nu_{\rm min} = \nu$
            \STATE {\bf end}
            \STATE {\bf if} {$|B - \bar{B}| < \tau_b$} {\bf then}
            \STATE ~~Break the loop
            \STATE {\bf end}
            \ENDFOR
    	\end{algorithmic}}}
    \end{algorithm}

    Based on Theorem 3, we determine the optimal quantization levels as summarized in {\bf Algorithm 1}. In this method, the optimal Lagrange multiplier $\nu^\star$ in {Theorem 3} is iteratively determined using a bisection search algorithm in conjunction with the Newton-Raphson method \cite{NR_method}, which is a representative root-finding algorithm, to determine $Q_i$ that satisfies $\nu^\star = h(Q_i ;w(a_i), \tilde{\mu})$ for the given $\nu^\star$. As proved in Appendix A, the function $h(Q_i;w(a_i), \tilde{\mu})$ is a strictly decreasing function of $Q_i$ when $\tilde{\mu} < \frac{3}{13}$. Therefore, the Newton-Raphson method yields a unique solution for $Q^\star_i$ that satisfies \eqref{eq:h_def} within a finite number of iterations. Note that if $\tilde{\mu}=0$, the problem $({\bf P}_4)$ becomes equivalent to the problem $({\bf P}_2)$ (i.e., $D(Q_i; \tilde{\mu}=0)=D_0/Q_i^2 $), and the optimal quantization level can be explicitly derived in the same manner as in Sec.~\ref{Sec:IAQ_error_free3}. Our modified water-filling method for determining the optimal quantization level in the problem $({\bf P}_4)$ is outlined in \textbf{Algorithm 1}. After completing \textbf{Algorithm 1}, we applied the method described in Sec.~\ref{Sec:IAQ_error_free3} to fully utilize the available bit budget.
    

    \vspace{1mm}
    {\bf Remark 2 (Complexity comparison of the modified methods):}
    As discussed in Sec.~\ref{Sec:IAQ_error_free3}, the optimal solution of $({\bf P}_4)$ is suboptimal for the original problem $({\bf P}_3)$.
    The computational complexity associated with solving $({\bf P}_4)$ is $\mathcal{O}(NS_{\rm max}T_{\rm max})$, where $T_{\rm max}$ denotes the number of iterations required by the Newton-Raphson method. This formulation yields a substantial advantage in comparison to the $\mathcal{O}(N\frac{\bar{B}}{P^2C})$ complexity of $({\bf P}_3)$, as it holds that $S_{\rm max}T_{\rm max} \leq N \leq \frac{\bar{B}}{P^2C}$. Consequently, this demonstrates that water-filling algorithm is more computationally efficient than incremental allocation method.

   \section{Simulation Results}\label{Sec:Simul}

   In this section, we evaluate the superiority of the proposed IAQ method through simulations. In these simulations, we consider the following tasks:
   \begin{itemize}
       \item {\bf Single-view image classification:} In this task, we consider single-view image classification using the CIFAR-100\footnote{The CIFAR-100 dataset consist of 50,000 training images and 10,000 test images.} \cite{CIFAR-100} dataset. 
           
       \item {\bf Multi-view image classification:} In this task, we consider multi-view image classification using the MIRO\footnote{The MIRO dataset comprises 12 classes, each containing 1600 images. These images are divided among 10 distinct objects, with 160 multi-view images per object. For the training dataset, 32 multi-view images per object are selected, while 128 images are reserved for testing. Consequently, the dataset includes 3,840 training images and 15,360 testing images.} \cite{MIRO}  and MVP-N\footnote{The MVP-N dataset consists of 44 classes with 1,760 training images and 7,040 test images. The testing images include one object per class, where each object is represented by 160 multi-view images.} \cite{MVP_N} datasets. We assume that four devices send the images for the same object, but from different views. The server receives a single-view image from each device and classifies the image individually. The final classification result is determined by applying a majority rule to the four individual outcomes.

   \end{itemize}

   All datasets are normalized to zero mean and unit variance. 
   The ViT encoder model on the device is DeiT-Tiny, while DeiT-Small is used on the server, with 4.4 times more parameters than the device model \cite{D_model}. The DeiT-Tiny and DeiT-Small models are pretrained on the ImageNet-1k dataset, consisting of 1 million images and 1,000 classes. The classifiers of both models consist of a single fully connected layer. The pretrained ViT encoder models and classifiers are also fine-tuned according to our datasets. During the fine-tuning, we use the cross-entropy loss for both single-view and multi-view classification tasks. 
   The Adam optimizer is applied with a learning rate of $0.0001$, and the batch size is set to $32$ across all datasets. The total number of epochs is set to $10$ for MVP-N, and $3$ for CIFAR-100 and MIRO. It should be noted that our fine-tuning process does not involve any communication process or channels. 
   The input image size for both device and server is $(3, 224, 224)$ across all datasets. Both models are characterized by an embedding dimension of 768, a patch size of 16, 12 encoder layers, 196 patches, and 12 attention heads (i.e., $P=16, L=12, N=196, H=12$). 
   The compression ratio is defined as the ratio of the compressed bit overhead to the original bit overhead (i.e., $\rho = \frac{P^2C\sum_{i=1}^{N}\log_2 Q_i}{8HWC}$). 
   For performance comparison, we consider the following methods:
    \begin{figure*}[t]
        \begin{minipage}{2\columnwidth}
            \centering
            \subfigure[$\tilde{\mu}=0$, MIRO]
            {\epsfig{file=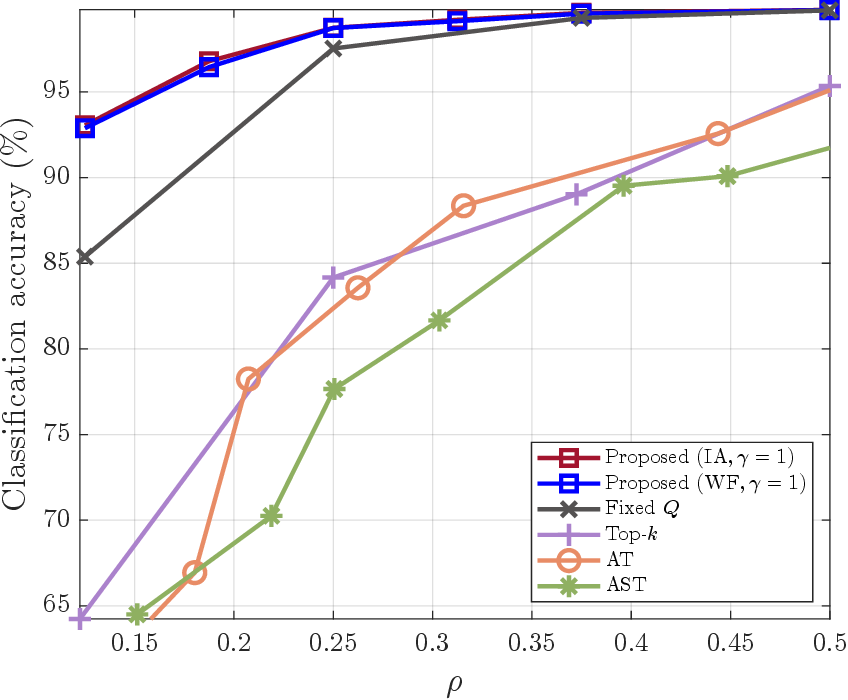, width=7cm}}
            \hspace{3mm}
            \subfigure[$\tilde{\mu}=0$, MVP-N]
		{\epsfig{file=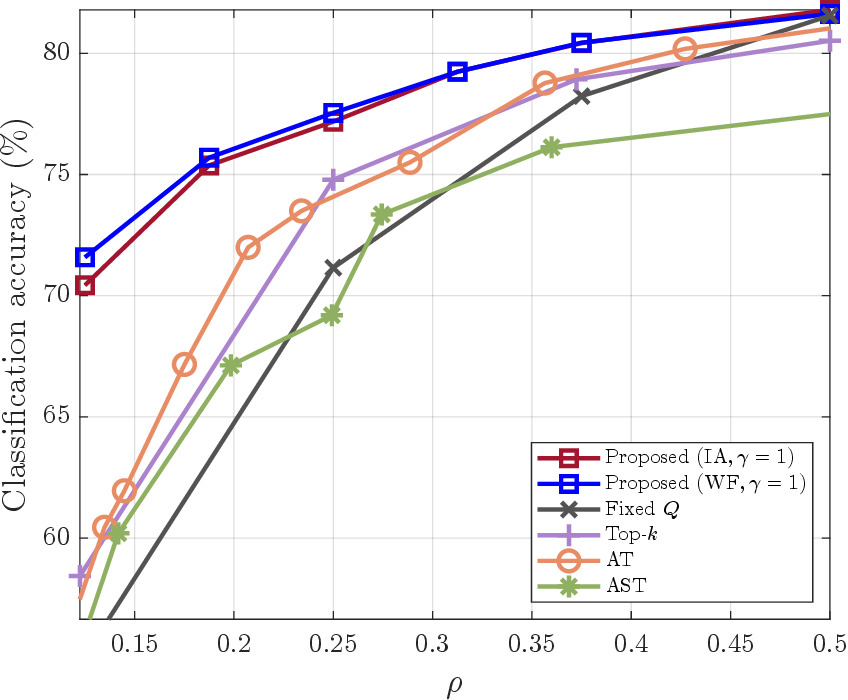, width=7cm}} 
            \captionof{figure}{
           Comparison of the classification accuracies of various quantization approaches for multi-view image classification tasks using the MIRO and MVP-N datasets.}
           \label{fig:EF_ACC}
        \end{minipage}
    \end{figure*}
   
    
   \begin{itemize}
    \item {\bf Proposed}: We consider four methods for the proposed IAQ framework: (i) {\bf Proposed (IA)}, which employs the incremental allocation method in Sec.~\ref{Sec:IAQ_error_free2} to solve the problem $({\bf P}_1)$; (ii) {\bf Proposed (WF)}, which employs the water-filling method in Sec.~\ref{Sec:IAQ_error_free3} to solve the problem $({\bf P}_2)$; (iii) {\bf Proposed (Modified IA)}, which employs the modified incremental allocation method in Sec.~\ref{Sec:IAQ_error3} to solve the problem $({\bf P}_3)$; and (iv) {\bf Proposed (Modified WF)}, which employs the modified water-filling method in Sec.~\ref{Sec:IAQ_error3} to solve the problem $({\bf P}_4)$. In all these methods, we set $S_{\rm max}=10$, $T_{\rm max}=5$, $M_{\rm max}=8$, and $Q_{\rm max}=256$. Therefore, as highlighted in {\bf Remark~1} and {\bf Remark~2}, the water-filling method is more computationally efficient than the incremental allocation method.
    A weight function $w(a_i)$ is chosen as 
    \begin{align}\label{eq:weight_fn}
        w(a_i) = \frac{1-d}{(a_{\rm max}-a_{\rm min})^\gamma}(a_i-a_{\rm min})^\gamma+d,
    \end{align}
    where $a_{\rm min} = \underset{i}{\rm min}(a_i)$, $a_{\rm max} = \underset{i}{\rm max}(a_i)$, $\gamma >0$ is a factor that determines the shape of the weight, and $d=10^{-7}$ is an arbitrary small value introduced to prevent the weight from becoming zero. 
    When $\gamma > 1$, the weight function becomes convex, whereas for $\gamma < 1$, it becomes concave. In the special case of $\gamma=1$, the weight function is linear with respect to the mean attention score.

    \item {\bf Fixed-$\bm Q$}: We consider the fixed-level quantization approach, which falls under the category of fixed-bit quantization. In this case, the compression ratio can be simplified to $\rho = \frac{NP^2C\log_2 Q_i}{8HWC}$, as $Q_i$ remains constant for all $i$.     

    \item {\bf Top-$\bm k$}: We consider the attention-aware patch selection based quantization approach in \cite{Transformer_att_1}. In this strategy, we allocate the highest quantization level, $Q_i = Q_{\rm max}$, to the top $k\%$ of patches based on their attention scores, while assigning the lowest quantization level, $Q_i = 1$, to the remaining patches. In this case, the compression ratio can be simplified to $\rho = \frac{NkP^2C\log_2 Q_{\rm max}}{100\times 8HWC}$. Therefore, the compression ratio can be determined based on the value of $k$.
    
    
    \item {\bf Attention threshold (AT)}: We consider the attention-aware patch selection based quantization approach in \cite{Transformer_att_1}. In this strategy, we allocate the highest quantization level, $Q_i = Q_{\rm max}$, to the patches whose attention scores exceed a certain threshold $\delta$, while assigning the lowest quantization level, $Q_i = 1$, to the remaining patches. 
    Since the precise relationship between $\delta$ and the compression ratio remains unclear, we numerically determine the optimal $\delta$ within the range of $0.001$ to $0.01$ for each simulation setting.

    

    
    \item {\bf Attention sum threshold (AST)}: We consider the attention-aware patch selection based quantization approach in \cite{Transformer_att_1}. In this strategy, we allocate the highest quantization level, $Q_i=Q_{\rm max}$, to patches in order of their attention scores, until the cumulative sum of these scores exceeds a specified threshold $\delta_{\rm sum}$. The remaining patches are assigned the lowest quantization level, $Q_i= 1$. 
    Since the precise relationship between $\delta_{\rm sum}$ and the compression ratio remains unclear, we numerically determine the optimal $\delta_{\rm sum}$ within the range of $0.3$ to $0.9$ for each simulation setting.



   \end{itemize}

    \subsection{Performance Evaluation under Error-Free Communication}

    Fig.~\ref{fig:EF_ACC} compares the classification accuracies of various quantization methods for the multi-view image classification tasks on the MIRO and MVP-N datasets. 
    Fig.~\ref{fig:EF_ACC} shows that the proposed methods achieve a higher classification accuracy compared to the existing quantization methods, particularly when the target communication overhead is low. 
    Although existing attention-aware quantization methods (i.e., {\bf Top-$\bm k$}, {\bf AT}, and {\bf AST}) consider mean attention scores, the proposed method offers additional performance gains over these methods by enabling optimal bit allocation beyond binary bit selection. 
    Additionally, {\bf Proposed (IA)} and {\bf Proposed (WF)} can control the communication overhead effectively. For instance, it achieves task performance at $\rho=0.1875$ and $\rho=0.3125$, which {\bf Fixed-$\bm Q$}, {\bf AT}, and {\bf AST} could not achieve. This demonstrates the flexibility of {\bf Proposed (IA)} and {\bf Proposed (WF)} in controlling communication overhead. 
    The performance of {\bf Proposed (IA)} and {\bf Proposed (WF)} is nearly identical. Since {\bf Proposed (WF)} requires lower computational complexity than {\bf Proposed (IA)}, {\bf Proposed (WF)} becomes a more appealing solution, even though it relies on some relaxations to determine the optimal bit allocation.

    
    \begin{figure}[t]
        \centering 
            {\epsfig{file=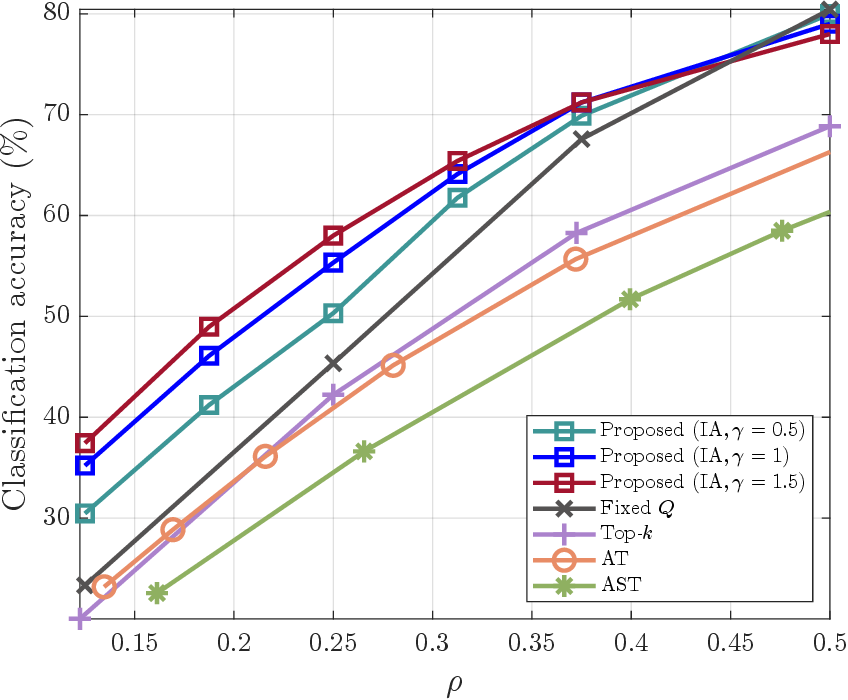, width=7cm}} 
        \caption{Comparison of the classification accuracies of proposed IAQ approaches for single-view image classification tasks on the CIFAR-100 dataset.}
        \label{fig:ada_mod_control_1}
    \end{figure}

    \begin{figure*}[t]
        \begin{minipage}{2\columnwidth}
            \centering
            \subfigure[$\rho = 0.125$]
            {\epsfig{file=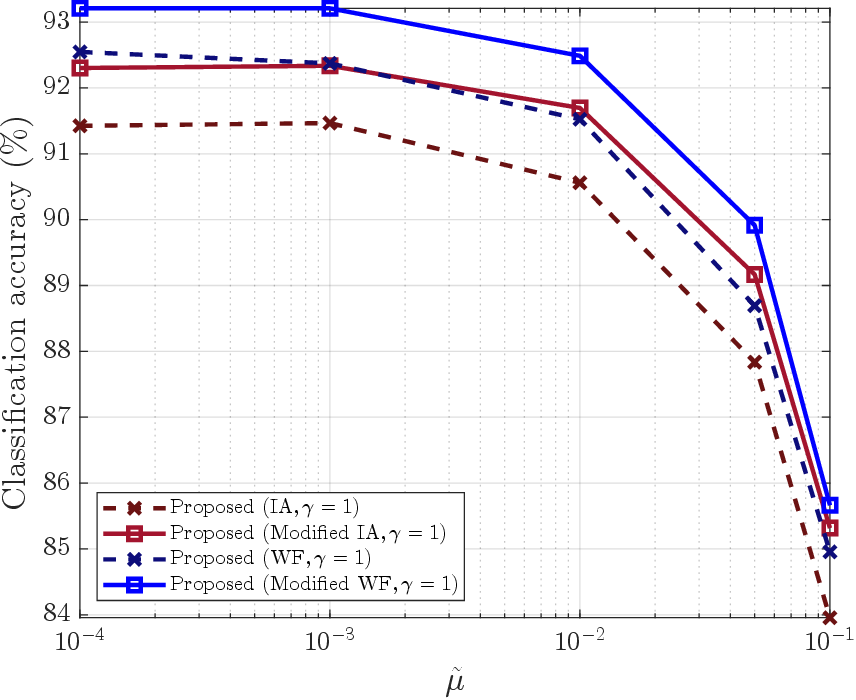, width=7cm}}
            \hspace{3mm}
            \subfigure[$\tilde{\mu} = 0.05$]
		{\epsfig{file=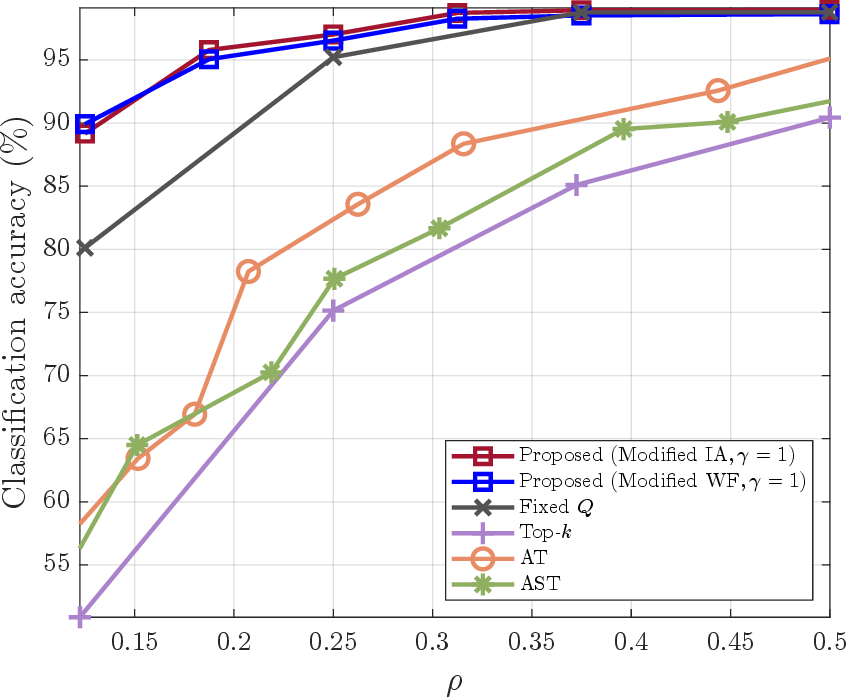, width=7cm}} 
            \captionof{figure}{
            Comparison of the classification accuracy and communication overhead of various quantization approaches under different channel error levels for a multi-view image classification task on the MIRO dataset.}
           \label{fig:Error_ACC_OH}
        \end{minipage}
    \end{figure*}
    

    \begin{figure*}[t]
        \centering 
            {\epsfig{file=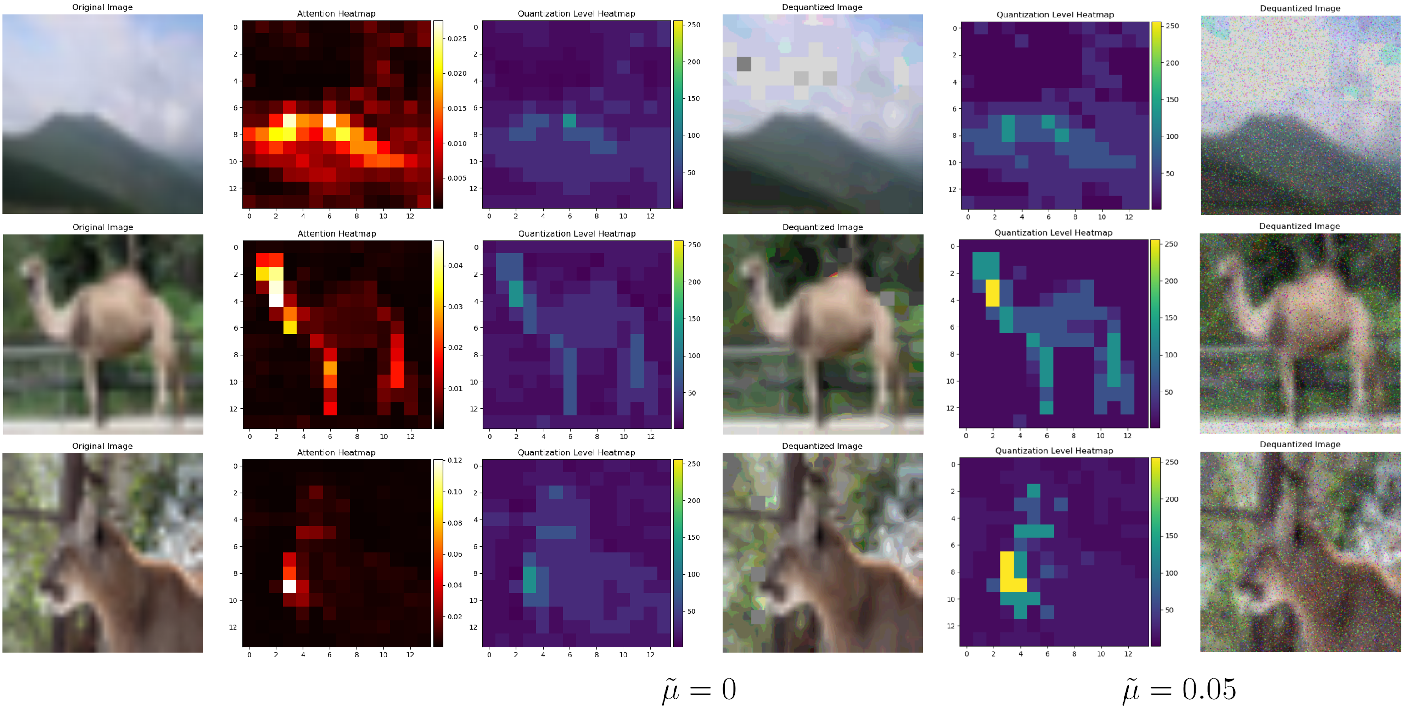, width=15.6cm}} 
        \caption{Visualization of the attention score map and the quantization level map of the proposed IAQ approach under both error-free and erroneous conditions at $\rho = 0.5$, for a single-view image classification task on the CIFAR-100 dataset.}
        \label{fig:Q_level_vis}
    \end{figure*} 

    Fig.~\ref{fig:ada_mod_control_1} compares the classification accuracies of various quantization methods for the single-view image classification task on the CIFAR-100 dataset. In this simulation, we also examine the effect of the weight function design by comparing the performance of {\bf Proposed (IA)} with different values of $\gamma$ in \eqref{eq:weight_fn}.
    Fig.~\ref{fig:ada_mod_control_1} shows that as $\gamma$ increases, the performance of {\bf Proposed (IA)} improves for $\rho \leq 0.375$, while lower $\gamma$ values yield better performance for $\rho > 0.375$.
    Specifically, an exponential-like weight function (i.e., $\gamma = 1.5$) is preferred when the target communication overhead is low, while a root-like weight function (i.e., $\gamma = 0.5$) is preferred when the communication overhead is high.
    This indicates that the performance of {\bf Proposed (IA)} can be maximized by designing the weight function according to the target communication overhead. Nevertheless,  {\bf Proposed (IA)} consistently outperforms {\bf  Fixed-$\bm Q$}, {\bf Top-k}, {\bf AT}, and {\bf AST} across all values of $\gamma$ for most cases. 
    These results demonstrate that the proposed method performs well with various choices of the weight function.




    \subsection{Performance Evaluation under BSC Models}

    Fig.~\ref{fig:Error_ACC_OH} compares the classification accuracy and communication overhead of various quantization methods under different bit-flip probabilities for the multi-view image classification task on the MIRO dataset.
    Fig.~\ref{fig:Error_ACC_OH}(a) shows that {\bf Proposed (Modified IA)} outperforms {\bf Proposed (IA)} and {\bf Proposed (Modified WF)} outperforms {\bf Proposed (WF)}. This indicates that our adaptive bit allocation by jointly considering both quantization and bit-flip errors offers robustness against communication errors. Furthermore, the superior performance of {\bf Proposed (WF)} compared to {\bf Proposed (IA)} and the better performance of {\bf Proposed (Modified WF)} over {\bf Proposed (Modified IA)} suggest that the low-complexity algorithms introduced in Sec.~\ref{Sec:IAQ_error_free}-B and Sec.~\ref{Sec:IAQ_error}-C exhibit even greater performance when $\rho=0.125$.
    Fig.~\ref{fig:Error_ACC_OH}(b) demonstrates that when $\tilde{\mu}=0.05$, the proposed IAQ approaches maintain strong task performance compared to the baseline methods, particularly when the target communication overhead is low.

    Fig.~\ref{fig:Q_level_vis} visualizes the attention score map and quantization level map of {\bf Proposed (WF)} under error-free communications (i.e. $\tilde{\mu} = 0$), and {\bf Proposed (Modified WF)} under BSCs with $\tilde{\mu} = 0.05$ at $\rho=0.5$, for the single-view image classification task on the CIFAR-100 dataset.
    Fig.~\ref{fig:Q_level_vis} shows that patches with higher attention scores are assigned higher quantization levels, both when $\tilde{\mu} = 0$ and $\tilde{\mu} = 0.05$, which is consistent with the results derived in \eqref{eq:opt_DM_5} and \eqref{eq:OBEA_DM_3}.
    Furthermore, a comparison between the dequantized images for different communication errors reveals that, as $\tilde{\mu}$ increases, the quantization level for background patches decreases, whereas it increases for the object regions.
    This allocation strategy helps ensure that crucial information for task performance remains robust under communication errors, while discarding less essential information.

   \section{Conclusion}
   In this paper, we have proposed a novel IAQ framework for training-free ViT-based semantic communications. The key idea behind our framework is to utilize a pretrained ViT model to quantify the importance levels of different image patches. Based on this idea, we have devised the optimal bit allocation for different image patches by formulating a weighted quantization error minimization problem and also by developing the allocation methods to solve this problem. We have also extended our framework to consider communication errors in realistic digital communication systems. By adopting the BSC modeling approach, we have successfully modified our importance-aware bit allocation methods. Our simulation results demonstrate that our IAQ framework significantly improves the performance-overhead tradeoff of both single-view and multi-view image classification tasks. 

   
   An important direction for future research is to develop a tractable framework that characterizes the relationship between attention scores and task performance. Expanding the IAQ approach to realistic wireless channels, particularly through integration with classical Shannon communications, offers promising prospects. Additionally, extending this framework to support multi-task, multi-modal semantic communications could enable concurrent task execution across devices.
   
    \appendices
    \section{Proof of Theorem 1 \& 3}\label{apdx:Q_opt}
    The Lagrangian function for \eqref{eq:OBEA_DM_21} is given by
    \begin{align}\label{eq:lagrange}
        &\mathcal{L}(\left\{Q_i\right\}^{N}_{i=1}, \nu, \left\{\lambda_i\right\}^{N}_{i=1}, \left\{\gamma_i\right\}^{N}_{i=1}) \nonumber\\
        &= \sum_{i=1}^{N}w(a_i) D(Q_i;\tilde{\mu}) + \nu\left(P^2C\sum_{i=1}^{N}\log_2 Q_i - \bar{B}\right)\nonumber\\
        &~~~+\sum_{i=1}^{N}\lambda_i(1-Q_i)+\sum_{i=1}^{N}\gamma_i(Q_i-Q_{\rm max}).
    \end{align}
    Then, the KKT conditions for \eqref{eq:lagrange} are represented by
    \begin{align}
         &1 \leq Q_i \leq Q_{\rm max}, \forall i \in \left\{1,\ldots,N\right\}, \label{KKT1} \\
        &\nu, \lambda_i, \gamma_i \geq 0, \forall i, \label{KKT2}\\
        &\lambda_i(1-Q_i) = 0, \forall i, \label{KKT3}\\
        &\gamma_i(Q_i-Q_{\rm max}) = 0, \forall i, \label{KKT4}\\
        &\nu\left(P^2C\sum_{i=1}^{N}\log_2 Q_i - \bar{B}\right) = 0, \label{KKT5}\\
        &  \gamma_i -\lambda_i + w(a_i)\frac{\partial D(Q_i;\tilde{\mu})}{\partial Q_i}+\frac{\nu P^2C}{Q_i\ln 2}=0, \forall i \label{KKT6}.
    \end{align}
    Under the condition in \eqref{KKT1}, we consider the following three cases:

    \begin{itemize}
        \item[(i)] {\bf Case 1} ($1<Q_i < Q_{\rm max}$): In this case, we have $\lambda_i = \gamma_i = 0$ from \eqref{KKT3} and \eqref{KKT4}. Therefore, the condition in \eqref{KKT6} can be rewritten as 
        \begin{align}\label{nu_equality}
          \nu   &= - \frac{w(a_i)\ln 2}{P^2C}Q_i \frac{\partial D(Q_i;\tilde{\mu})}{\partial Q_i}  = h(Q_i;w(a_i),\tilde{\mu}),
        \end{align}
        where $h(Q_i;w(a_i),\tilde{\mu})$ is given in \eqref{eq:h_def}. In this case, the value of $Q_i$ can be determined by finding the solution of the equation in \eqref{nu_equality} for a given value of $\nu$.

        \item[(ii)] {\bf Case 2} ($Q_i=1$): In this case, we have $\lambda_i \geq 0$ and  $\gamma_i = 0$ from \eqref{KKT2}--\eqref{KKT4}. Therefore, the condition in \eqref{KKT6} implies that 
        \begin{align}
          w(a_i)\frac{\partial D(Q_i;\tilde{\mu})}{\partial Q_i} \Big|_{Q_i=1}+\frac{\nu P^2C}{\ln 2} \geq 0.
        \end{align}
        This is equivalent to $\nu \geq h(1;w(a_i),\tilde{\mu})$.

        \item[(iii)] {\bf Case 3} ($Q_i=Q_{\rm max}$): In this case, we have $\lambda_i = 0$ and  $\gamma_i \geq 0$ from \eqref{KKT2}--\eqref{KKT4}. Therefore, the condition in \eqref{KKT6} implies that 
        \begin{align}
          w(a_i)\frac{\partial D(Q_i;\tilde{\mu})}{\partial Q_i} \Big|_{Q_i=Q_{\rm max}}+\frac{\nu P^2C}{Q_{\rm max} \ln 2} \leq 0.
        \end{align}
        This is equivalent to $\nu \leq h(Q_{\rm max};w(a_i),\tilde{\mu})$.
    \end{itemize}

    Now, we prove that $h(Q_i;w(a_i), \tilde{\mu})$ is a strictly decreasing function of $Q_i$. 
    The first derivative of $h(Q_i;w(a_i),\tilde{\mu})$ with respect to $Q_i$ is given by 
    \begin{align}\label{h_diff}
            \frac{\partial h(Q_i;w_i, \tilde{\mu})}{\partial Q_i}  =&  \frac{w (a_i) D_0 \ln 2}{P^2C(1-\tilde{\mu})}  Q_i^{x-1} g(Q_i;\tilde{\mu}),
        \end{align}
        where $x=\log_2 (\frac{1-\tilde{\mu}}{4})$ and $g(Q_i;\tilde{\mu}) $ is defined as
        \begin{align}
        g(Q_i;\tilde{\mu}) =& -\frac{4}{3}\tilde{\mu}(x+2)^2Q_i^2 - 4\tilde{\mu}(x+1)^2Q_i \nonumber \\
        &- x^2\left(1-\frac{16}{3}\tilde{\mu}\right).
    \end{align}
    The function $g(Q_i;\tilde{\mu})$ satisfy the following inequality:
    \begin{align}
        g(Q_i;\tilde{\mu}) &\overset{(a)}{\leq} -4\tilde{\mu}Q_i-4\left(1-\frac{16}{3}\tilde{\mu}\right) \nonumber \\
            &\overset{(b)}{\leq} -4\tilde{\mu}-4\left(1-\frac{16}{3}\tilde{\mu}\right) \overset{(c)}{<} 0,
        \end{align}
        where $(a)$ follows from the condition $\tilde{\mu} \geq 0$ and $x \leq -2$, $(b)$ is based on the fact that $Q_i \geq 1$, and $(c)$ holds because $\tilde{\mu} <\frac{3}{13}$.
        Therefore, $\frac{\partial h(Q_i;w_i, \tilde{\mu})}{\partial Q_i} < 0$, indicating that $h(Q_i;w(a_i), \tilde{\mu})$ is a strictly decreasing function of $Q_i$. This also implies that $Q_i = h^{-1}(\nu;w(a_i),\tilde{\mu})$ exists and is a strictly decreasing function of $\nu$.

    By combining all the above results, the optimal quantization level $Q_i^\star$ that satisfies the KKT conditions is given by 
    \begin{align}
        Q_i^\star = 
        \begin{cases}\label{eq:final_q_level}
          1, &\nu^\star \geq h(1;w(a_i), \tilde{\mu}), \\ 
          Q_{\rm max}, &\nu^\star \leq h(Q_{\rm max};w(a_i), \tilde{\mu}), \\
          h^{-1}(\nu^\star;w(a_i),\tilde{\mu}), &\text{otherwise},
        \end{cases}
    \end{align}
    where $\nu^\star$ is set to satisfy the equality of $P^2C\sum_{i=1}^{N}\log_2 Q_i^\star = \bar{B}$.
    The above expression can be rewritten as given in \eqref{eq:OBEA_DM_3}, which completes the proof for Theorem 3.

    We also prove {Theorem 1} by setting $\tilde{\mu} = 0$ in \eqref{eq:final_q_level}. If $\tilde{\mu} = 0$, the function $h(Q_i; w(a_i), \tilde{\mu})$ in \eqref{eq:h_def} is expressed as 
    \begin{align}
        h(Q_i; w(a_i), 0) =2\frac{w (a_i) D_0 \ln 2}{P^2C}Q_i^{-2}.
    \end{align}
    Consequently, we have  
    \begin{align}
        h^{-1}(\nu;w(a_i), 0) = \sqrt{\frac{w(a_i)\ln 2(u_{\rm max}-u_{\rm min})^2}{2\nu}}.
    \end{align}
    Therefore, the optimal quantization level $Q_i^\star$ is rewritten as
    \begin{align}
        Q_i^\star = 
        \begin{cases}\label{eq:final_q_level_2}
          1, &\nu^\star \geq h(1;w(a_i), 0), \\
          Q_{\rm max}, &\nu^\star \leq h(Q_{\rm max};w(a_i), 0), \\
          \sqrt{\frac{w(a_i)\ln 2(u_{\rm max}-u_{\rm min})^2}{2\nu^\star}}, &\text{otherwise}.
        \end{cases}
    \end{align}
       The above expression can be rewritten as given in \eqref{eq:opt_DM_5}, which completes the proof for Theorem 1.


    \section{Proof of Theorem 2}\label{apdx:the2}
    \begin{figure*}
    \begin{align}\label{eq:OBEA_DM_1}
        \sum_{k=1}^{P^2C}\mathbb{E}_{{\bf b}_{ik},\hat{\bf b}_{ik}}[|u_{ik}-\hat{u}_{ik}|^2] 
        &\overset{(a)}{\leq} \sum_{k=1}^{P^2C}\sum_{l_1,\ldots,l_{M_i} \in \{0,1\}}^{}\sum_{\hat{l}_1,\ldots,\hat{l}_{M_i} \in \{0,1\}}^{}\prod_{r=1}^{M_i}\mathbb{P}[b_{{ik},r}=l_r]\mathbb{P}[\hat{b}_{{ik},r}=\hat{l}_r|b_{{ik},r}=l_r]\cdot |u_{ik}-\hat{u}_{ik}|^2 \nonumber \\
        &\overset{(b)}{\leq} \sum_{k=1}^{P^2C}\sum_{l_1,\ldots,l_{M_i} \in \{0,1\}}^{}\sum_{\hat{l}_1,\ldots,\hat{l}_{M_i} \in \{0,1\}}^{}\left(\frac{1}{2}\right)^{M_i}\prod_{r=1}^{M_i}\mathbb{P}[\hat{b}_{{ik},r}=\hat{l}_r|b_{{ik},r}=l_r]\cdot |u_{ik}-\hat{u}_{ik}|^2 \nonumber \\
        &\overset{(c)}{\leq} \sum_{k=1}^{P^2C}\Bigg[\underbrace{(1-\tilde{\mu})^{M_i}\left(\frac{\Delta_i}{2}\right)^2}_{E_0}+\underbrace{(1-\tilde{\mu})^{M_i-1}\tilde{\mu}\left(\frac{\Delta_i}{2}\right)^2\sum_{t=1}^{M_i}\left(2^{M_i+1}\left(\frac{1}{2}\right)^t+1\right)^2}_{E_1}\Bigg] \nonumber\\
        &\overset{(d)}{=} \sum_{k=1}^{P^2C}\frac{(u_{{\rm max}}-u_{{\rm min}})^2}{4(1-\tilde{\mu})}Q_{i}^{\log_2 (\frac{1-\tilde{\mu}}{4})}\left\{\frac{4}{3}\tilde{\mu} Q_i^2+4 \tilde{\mu} Q_i + \left(1-\frac{16}{3}\tilde{\mu} \right) \right\} \nonumber\\
        &=\frac{D_0}{(1-\tilde{\mu})}Q_{i}^{\log_2 (\frac{1-\tilde{\mu}}{4})}\left\{\frac{4}{3}\tilde{\mu} Q_i^2+4 \tilde{\mu} Q_i + \left(1-\frac{16}{3}\tilde{\mu} \right) \right\}=D(Q_i;\tilde{\mu}). 
    \end{align}
    \hrulefill	
    \end{figure*} 

    The upper bound of the expected distortion for the $i$-th patch is presented in \eqref{eq:OBEA_DM_1}. Specifically, $(a)$ follows from the property of memoryless BSCs, $(b)$ follows from the assumption of equiprobable bit outputs, $(c)$ denotes the exact upper bound of expected distortion of the $i$-th patch under {\bf Assumption 1}, where $E_m$ denoting the upper bound of the mean squared error (MSE) between $u_{ik}$ and $\hat{u}_{ik}$ when ${\bf b}_{ik}$ undergoes $m \in \{0,1\}$ bit errors, and $(d)$ follows from the fact that $Q_i=2^{M_i}$.


    \section{Proof of Proposition 1}\label{apdx:pro1}
    \begin{figure*}
    \begin{align}\label{eq:D_one_diff}
        \frac{\partial D(2^{M_i};\tilde{\mu})}{\partial M_i} = \frac{K\ln 2}{1-\tilde{\mu}}2^{xM_i}\bigg(\underbrace{\frac{4}{3}\tilde{\mu}(x+2)\cdot2^{2M_i}+4\tilde{\mu}(x+1)\cdot2^{M_i}+x\left(1-\frac{16}{3}\tilde{\mu}\right)}_{\triangleq D_{1,i}}\bigg).
    \end{align}
    \hrulefill	
    \end{figure*} 
    By utilizing the distortion in \eqref{eq:OBEA_DM_1}, the first derivative of $D(2^{M_i};\tilde{\mu})$ can be obtained as \eqref{eq:D_one_diff} where $x=\log_2(\frac{1-\tilde{\mu}}{4})$. 
    The upper bound of $D_{1,i}$ in \eqref{eq:D_one_diff} can be determined as follows:    
    \begin{align}\label{eq:D_1}
        D_{1,i} &\overset{(a)}{\leq} -4\tilde{\mu}\cdot 2^{M_i}-2\left(1-\frac{16}{3}\tilde{\mu}\right) \nonumber \\
        &\overset{(b)}{\leq} -4\tilde{\mu}-2\left(1-\frac{16}{3}\tilde{\mu}\right)  =  \frac{20}{3}\tilde{\mu} - 2,
    \end{align}
    where $(a)$ follows from the condition $\tilde{\mu} \geq 0$ and $x \leq -2$, and $(b)$ is based on the fact that $M_i \geq 0$.
    Therefore, if $\tilde{\mu} \leq 0.3$, we have $\frac{\partial D(2^{M_i};\tilde{\mu})}{\partial M_i} \leq 0$, implying that $D(2^{M_i};\tilde{\mu})$ is a monotonically decreasing function of $M_i$.

    \begin{figure*}
    \begin{align}\label{eq:D_two_diff}
        \frac{\partial^2 D(2^{M_i};\tilde{\mu})}{\partial M_i^2} = \frac{K(\ln 2)^2}{1-\tilde{\mu}}2^{xM_i}\bigg(\underbrace{\frac{4}{3}\tilde{\mu}(x+2)^2\cdot2^{2M_i}+4\tilde{\mu}(x+1)^2\cdot2^{M_i}+x^2\left(1-\frac{16}{3}\tilde{\mu}\right)}_{\triangleq D_{2,i}}\bigg).
    \end{align}
    \hrulefill	
    \end{figure*}  
    Similarly, the second derivative of $D(2^{M_i};\tilde{\mu})$ is expressed in \eqref{eq:D_two_diff}. 
    The lower bound of $D_{2,i}$ in \eqref{eq:D_two_diff} can be determined as follows: 
    \begin{align}\label{eq:D_2}
        D_{2,i} &\overset{(a)}{\geq} 4\tilde{\mu}2^{M_i} + 4\left(1-\frac{16}{3}\tilde{\mu}\right) \nonumber \\
        &\overset{(b)}{\geq} 4\tilde{\mu} + 4\left(1-\frac{16}{3}\tilde{\mu}\right)  = - \frac{52}{3}\tilde{\mu} + 4.
    \end{align}
    where $(a)$ follows from the condition $\tilde{\mu} \geq 0$ and $x \leq -2$, and $(b)$ is based on the fact that $M_i \geq 0$. 
    Therefore, if $\tilde{\mu} \leq \frac{3}{13}$, we have $\frac{\partial^2 D(2^{M_i};\tilde{\mu})}{\partial M_i^2} \geq 0$, implying that $D(2^{M_i};\tilde{\mu})$ is convex with respect to $M_i$. 
    Aggregating the above results implies that if  $\tilde{\mu} \leq \frac{3}{13}$, the distortion function $D(2^{M_i};\tilde{\mu})$ is convex and a monotonically decreasing function of $M_i$.


    \section{Proof of Proposition 2}\label{apdx:pro2}
    Let $D(2^{M_i};\tilde{\mu})=J_{\tilde{\mu}}(M_i)$ and $M_i=\log_2 Q_i=U(Q_i)$, then the second derivative of $D(Q_i;\tilde{\mu})$  with respect to $Q_i$ is derived as
    \begin{align}\label{eq:D_3}
        \frac{\partial^2 D(Q_i;\tilde{\mu})}{\partial Q_i^2} &= \frac{\partial^2 J_{\tilde{\mu}}(U(Q_i))}{\partial U(Q_i)^2}\left\{\frac{\partial U(Q_i)}{\partial Q_i}\right\}^2 \nonumber \\
        &~~~~+\frac{\partial J_{\tilde{\mu}}(U(Q_i))}{\partial U(Q_i)}\frac{\partial^2 U(Q_i)}{\partial Q_i^2} \nonumber \\
        &= \frac{1}{Q_i^2}\left(\frac{\partial^2 J_{\tilde{\mu}}(M_i)}{\partial M_i^2}I^2-\frac{\partial J_{\tilde{\mu}}(M_i)}{\partial M_i}I\right),
    \end{align}
    where $I=\frac{1}{\ln 2}$. If $\tilde{\mu} \leq \frac{3}{13}$, we have $\frac{\partial^2 J_{\tilde{\mu}}(M_i)}{\partial M_i^2} \geq 0$ and $\frac{\partial J_{\tilde{\mu}}(M_i)}{\partial M_i} \leq 0$ as shown in Appendix C. This directly implies that if $\tilde{\mu} \leq \frac{3}{13}$ and $Q_i \geq 1$, we have $\frac{\partial^2 D(Q_i;\tilde{\mu})}{\partial Q_i^2} \geq 0$. This completes the proof.

\bibliographystyle{IEEEtran}
\bibliography{Reference}
\end{document}